\documentclass[a4paper,12pt]{article} 

\newcommand{\beq}{\begin{equation}} 
\newcommand{\eeq}{\end{equation}} 
\newcommand{\bea}{\begin{eqnarray} }
\newcommand{\eea}{\end{eqnarray}} 
\newcommand{\bsub}{\begin{subequations}} 
\newcommand{\esub}{\end{subequations} \noindent} 
 
\makeatletter
\newtoks\@stequation

\def\subequations{\refstepcounter{equation}%
  \edef\@savedequation{\the\c@equation}%
  \@stequation=\expandafter{\theequation}
  \edef\@savedtheequation{\the\@stequation}
  \edef\oldtheequation{\theequation}%
  \setcounter{equation}{0}%
  \def\theequation{\oldtheequation\alph{equation}}}

\def\endsubequations{%
  \ifnum\c@equation < 2 \@warning{Only \the\c@equation\space subequation
    used in equation \@savedequation}\fi
  \setcounter{equation}{\@savedequation}%
  \@stequation=\expandafter{\@savedtheequation}%
  \edef\theequation{\the\@stequation}%
  \global\@ignoretrue}

\def\eqnarray{\stepcounter{equation}\let\@currentlabel\theequation
\global\@eqnswtrue\m@th
\global\@eqcnt\z@\tabskip\@centering\let\\\@eqncr
$$\halign to\displaywidth\bgroup\@eqnsel\hskip\@centering
     $\displaystyle\tabskip\z@{##}$&\global\@eqcnt\@ne
      \hfil$\;{##}\;$\hfil
     &\global\@eqcnt\tw@ $\displaystyle\tabskip\z@{##}$\hfil
   \tabskip\@centering&\llap{##}\tabskip\z@\cr}

\makeatother

\begin{document} 
\thispagestyle{empty} 
\vspace*{-15mm} 
\baselineskip 10pt 
\begin{flushright} 
\begin{tabular}{l} 
{\bf December 2001}\\ 
{\bf KEK-TH-793}\\ 
{\bf hep-ph/0112102} 
\end{tabular} 
\end{flushright} 
\baselineskip 24pt 
\vglue 10mm 
\begin{center} 
{\Large\bf 
 Comment on the sign  \\ 
 of the pseudoscalar pole contribution \\ 
 to the muon $g-2$ 
}
\vspace{8mm} 

\baselineskip 18pt 
\def\thefootnote{\fnsymbol{footnote}} 
\setcounter{footnote}{0} 
{\bf 
 Masashi Hayakawa 
  \footnote{e-mail address : haya@post.kek.jp} 
 and 
 Toichiro Kinoshita 
  \footnote{e-mail address : tk@mail.lns.cornell.edu} 
} 
\vspace{5mm} 

$^*${\it Theory Division, KEK, Tsukuba, Ibaraki 305-0801, Japan} 
\\ 
$^\dagger${\it Newman Laboratory, 
               Cornell University, Ithaca, New York 14853} 

\vspace{10mm} 
\end{center} 

\addtocounter{footnote}{-2} 

\begin{center} 
{\bf Abstract}\\[7mm] 
\begin{minipage}{12cm} 
\baselineskip 16pt 
\noindent 
 We correct the error in 
the sign of the pseudoscalar pole contribution 
to the muon $g-2$, 
which dominates the ${\cal O}(\alpha^3)$ 
hadronic light-by-light scattering effect. 
 The error originates from our oversight of a feature of  
the algebraic manipulation program FORM 
which defines the $\epsilon$-tensor in such a way that 
it satisfies the relation  
$\epsilon_{\mu_1 \mu_2 \mu_3 \mu_4} 
 \epsilon_{\nu_1 \nu_2 \nu_3 \nu_4} 
 \eta^{\mu_1 \nu_1} \eta^{\mu_2 \nu_2} 
 \eta^{\mu_3 \nu_3} \eta^{\mu_4 \nu_4} = 24$, 
irrespective of space-time metric. 
 To circumvent this problem, we replaced the product  
$\epsilon_{\mu_1 \mu_2 \mu_3 \mu_4} 
 \epsilon_{\nu_1 \nu_2 \nu_3 \nu_4}$ 
by 
$ - \eta_{\mu_1 \nu_1} \eta_{\mu_2 \nu_2} 
    \eta_{\mu_3 \nu_3} \eta_{\mu_4 \nu_4} \pm \cdots$ 
in the FORM-formatted program, 
and obtained a positive value for the pseudoscalar pole contribution, 
in agreement with the recent result obtained by Knecht {\it et al}. 

\end{minipage} 
\end{center} 
%
\newpage 
\baselineskip 18pt 
 In this brief article 
we report  
the result of our reexamination 
of the pseudoscalar pole contribution to the muon $g-2$. 
 In the previous studies 
\cite{Hayakawa:1995ps,Bijnens:1995cc}, 
the pseudoscalar pole contribution 
had been noted to be the dominant term 
of the ${\cal O}(\alpha^3)$ 
hadronic light-by-light scattering contribution to 
the muon $g-2$, $a_\mu \equiv (g_\mu - 2)/ 2$. 
 In view of the expected accuracy 
of the new muon $g-2$ measurement 
at the Brookhaven National Laboratory (BNL) 
($\Delta a_\mu({\rm exp}) = 4 \times 10^{-10}$), 
the present authors and Bijnens {\it et al.} 
have examined this contribution carefully 
\cite{Bijnens:1995cc,Hayakawa:1997rq}, 
taking account of the experimental data 
obtained for the $P \gamma^* \gamma$-vertex 
($P = \pi^0, \eta, \eta^\prime$) at CLEO \cite{Gronberg:1997fj}.  
 The primary purpose of this paper is to examine the sign 
of this contribution in light of the two recent papers 
\cite{Knecht:2001qf,Knecht:2001qg}. 
 We therefore concentrate on the $\pi^0$ pole contribution 
which gives the value  
\begin{equation} 
 a_\mu(\pi^0) = -55.60~(3) \times 10^{-11} \, , 
  \label{eq:pi0} 
\end{equation} 
in the naive vector meson dominance (nVMD) model 
\cite{Hayakawa:1995ps}. 
 This model simply modifies the Wess-Zumino term 
for the $\pi^0 \gamma \gamma$-vertex 
by attaching $\rho$-meson propagators 
which carry photon momenta $k_1, k_2$, 
\begin{equation} 
 - i \frac{\alpha}{\pi f_\pi} 
 \epsilon_{\mu\nu\alpha\beta} k_1^\alpha k_2^\beta \, 
 \frac{M_\rho^2}{M_\rho^2 - k_1^2} 
 \frac{M_\rho^2}{M_\rho^2 - k_2^2} 
   \, . 
\end{equation} 
 Both Bijnens {\it et al.}~\cite{Bijnens:1995cc} 
and Bartos {\it et al.}~\cite{Bartos:2001pg} obtained 
the negative value for the pseudoscalar pole contribution 
independently of our result. 
 Taking this value into account 
as a part of the standard model prediction, 
it is found that the current measurement of $g-2$ 
indicates $2.6~\sigma$ discrepancy from the prediction
of the standard model 
\cite{Brown:2001mg,Czarnecki:2001pv}. 
 
 However, 
the recent papers \cite{Knecht:2001qf,Knecht:2001qg} 
pointed out that the pseudoscalar pole contribution 
is positive, which is opposite to the sign 
of (\ref{eq:pi0}). 
 If this is true, it will reduce the deviation of 
the current BNL measurement 
from the standard model considerably. 
 In view of these papers \cite{Knecht:2001qf,Knecht:2001qg}, 
and of the significance of 
the ${\cal O}(\alpha^3)$ 
hadronic light-by-light scattering contribution 
in interpreting the current measured value of $a_\mu$, 
we decided to reinvestigate the $\pi^0$ pole contribution. 
 
 In the following, 
we summarize the results of our investigation 
to show various points we have scrutinized closely. 
 A full account is being prepared in a separate paper \cite{HK} . 
\vspace{0.5cm} 
 
 The first phase of examination consists of 
the following four steps of the original calculation 
\cite{Hayakawa:1995ps}: \\ 
(1) We obtained the expression for 
the contribution to the $\bar{\mu} \mu \gamma$-vertex 
function $\Gamma^\nu(p_I, p_F)$ 
($p_I \equiv p-q/2$ and $p_F \equiv p+q/2$ 
are the initial and final momenta of 
the muon) 
by direct application of Feynman rules. \\ 
(2) We evaluated the trace of the $\gamma$ matrices 
using the algebraic manipulation program FORM 
to extract $a_\mu$ from $\Gamma^\nu(p_I, p_F)$ 
by means of the magnetic moment projection 
\footnote{ 
 We correct the typo-error 
for the expression for $P_\nu(p, q)$ 
of eq.~(3.24) in Ref.~\cite{Hayakawa:1995ps}, 
which does not affect the result obtained there. 
}, 
\begin{eqnarray} 
 a_\mu &=& 
 \lim_{q^2 \rightarrow 0} 
 \lim_{\{ p \cdot q \rightarrow 0;\, p^2 + q^2 /4 \rightarrow m^2 \}} 
  {\rm Tr} 
  \left( 
   P_\nu(p, q) \Gamma^\nu(p_I, p_F) 
  \right) \, , 
   \nonumber \\ 
 P^\nu(p, q) &\equiv& 
 \displaystyle{ 
  \frac{m}{16\, p^4 q^2} 
  \left( 
   {p \hspace{-5.7pt}/\hspace{1.0pt}} - 
   \frac{{q \hspace{-5.8pt}/\hspace{1.0pt}}}{2} + m 
  \right) 
  \left( 
   \left( 
       \gamma^\nu {q \hspace{-5.8pt}/\hspace{1.0pt}} 
     - {q \hspace{-5.8pt}/\hspace{1.0pt}} \gamma^\nu 
   \right) p^2 
   - 3\, q^2 p^\nu 
  \right) 
 } \nonumber \\ 
 && \qquad \quad 
 \displaystyle{ 
  \times 
  \left( 
   {p \hspace{-5.7pt}/\hspace{1.0pt}} + 
   \frac{{q \hspace{-5.8pt}/\hspace{1.0pt}}}{2} + m 
  \right) \, , 
 } 
  \label{eq:projection} 
\end{eqnarray} 
where $q$ is the incoming photon momentum. \\ 
(3) The result of Eq.~(\ref{eq:projection}) is plugged 
into a FORTRAN program written in 
the formalism developed 
for the numerical evaluation of Feynman diagrams 
\cite{Cvitanovic:1974uf}. \\ 
(4) The numerical evaluation of $a_\mu$ is 
carried out with the help of the Monte Carlo integration routine VEGAS 
\cite{Lepage}. 

 In addition we confirmed 
by hand calculation and by MATHEMATICA 
that the projection operator 
in Eq.~(\ref{eq:projection}) works correctly, 
by extracting the anomalous magnetic moment 
$a_\mu = F_M(0)$ from 
\begin{eqnarray} 
 \Gamma^\nu(p_I, p_F) 
 &=& 
 \left( 
    F_E(q^2) \gamma^\nu 
  + F_M(q^2) \frac{1}{2 m} i \sigma^{\nu\lambda} q_\lambda 
 \right) \, , 
  \label{eq:form_factor} 
\end{eqnarray} 
with 
\begin{equation} 
 \sigma^{\mu\nu} 
  = \frac{i}{2} 
    \left[ 
     \gamma^\mu, \gamma^\nu 
    \right] \, . 
\end{equation} 

 After going through these steps \\ 
\\ 
(I) we rederived the value given in Eq.~(\ref{eq:pi0}) 
for the $\pi^0$ pole contribution in the nVMD model. 
\vspace{0.5cm} 
 
 The next phase was to derive 
the result obtained in Ref.~\cite{Knecht:2001qf}. 

 We first assumed that 
the expression given for $a_\mu$ in Eqs.~(3.4) and (3.5) 
of Ref.~\cite{Knecht:2001qf} is correct 
and see if the numerical evaluation of them 
gives the value of our Eq.~(\ref{eq:pi0}) 
but with opposite sign, 
as claimed in Ref.~\cite{Knecht:2001qf}. 
 We translated Eqs.~(3.4) and (3.5)  
into an expression 
suitable for the numerical evaluation 
according to the formalism of Ref.~\cite{Cvitanovic:1974uf}. 
 By carrying out 
the five-dimensional integration over the Feynman parameters 
with use of VEGAS, 
we obtain the value in Eq.~(\ref{eq:pi0}) 
but with an opposite sign. 
 This test not only confirms the result of 
Ref.~\cite{Knecht:2001qf} but also gives an evidence that \\ 
\\ 
(II) both Ref.~\cite{Knecht:2001qf} and our work performed 
the loop integration part correctly. 
 The disagreement in sign 
must therefore come from an earlier stage.  
\vspace{0.5cm} 

 We therefore switched our attention to the task 
of projecting $a_\mu$ out from $\Gamma^\nu(p_I, p_F)$ 
by means of algebraic programs. 
 Using the algebraic manipulation program FORM 
we pursued the steps described  
in Ref.~\cite{Knecht:2001qf} in detail 
to see whether we can reproduce their Eqs.~(3.4) and (3.5). 
 We found that \\ 
\\ 
(III) the trace operation by FORM yielded the sign {\it opposite}  
to that of $T_{1,2}(q_1, q_2; p)$ 
in Eq.~(3.5) of Ref.~\cite{Knecht:2001qf}. 
\vspace{0.5cm} 
 
 We thus recognized the following two possibilities 
as most likely: \\ 
(a) Ref.~\cite{Knecht:2001qf} made a mistake 
in picking up the sign of the trace of the $\gamma$ matrices. \\ 
(b) We made a mistake at the stage of 
picking up the sign of the trace of the $\gamma$ matrices 
{\it systematically}.  
 By systematically we mean that 
we failed to identify the sign irrespective 
of what projection operator was used to extract $a_\mu$; 
we derived the value of Eq.~(\ref{eq:pi0}) by 
both our own projection operator (\ref{eq:projection}) 
and those given in Eqs.~(2.9) - (2.11) of  Ref. \cite{Knecht:2001qf}. 
 All projectors gave the same result. 

 In examining the possibility (b), 
we noticed one crucial difference between 
Ref.~\cite{Knecht:2001qf}, which leads to the positive value, 
and ours, which leads to the negative value; 
while Ref.~\cite{Knecht:2001qf} 
used the algebraic manipulation program REDUCE 
to perform the trace calculation of the $\gamma$ matrices, 
we used FORM instead. 
 Recall that 
we have used FORM even 
for examining the results of Ref.~\cite{Knecht:2001qf}. 
 
 Thus we decided to check whether we handled FORM properly. 
 The program FORM had been used 
successfully to calculate the QED corrections to the $g-2$ of the muon 
and the electron 
as well as other observables by one of the present authors. 
 However, this does not guarantee that we deal correctly
with the $\epsilon$-tensor, 
the central object of our study of the pseudoscalar contribution. 
\

 A simple test of this question is to see 
if our naive use of the FORM declaration 
(FixIndex $1:-1, 2-1, 3:-1;$) 
works successfully to verify the identity 
\begin{equation} 
 \epsilon_{\mu_1 \mu_2 \mu_3 \mu_4} 
 \epsilon_{\nu_1 \nu_2 \nu_3 \nu_4} 
 \eta^{\mu_1 \nu_1} \eta^{\mu_2 \nu_2} 
 \eta^{\mu_3 \nu_3} \eta^{\mu_4 \nu_4} = -24 \, , 
\end{equation} 
which should hold in Minkowski space-time. 
 Unfortunately, the result turned out to be $+24$. 
 This means that this simple declaration does not work for 
the $\epsilon$-tensor in Minkowski space-time 
\footnote{ 
We thank J.~Vermasseren for explaining how to use 
the $\epsilon$-tensor for the Minkowski metric.
}. 
 In order to circumvent this difficulty 
we avoided the direct use of $\epsilon$-tensor 
and repeated the trace calculation 
using the right-hand side of the identity in Minkowski space-time: 
\begin{eqnarray} 
 \epsilon_{\mu_1 \mu_2 \mu_3 \mu_4} 
 \epsilon_{\nu_1 \nu_2 \nu_3 \nu_4} 
 &=& 
 \displaystyle{ 
  - 
  \left[ 
   \eta_{\mu_1 \nu_1} \eta_{\mu_2 \nu_2} 
    \eta_{\mu_3 \nu_3} \eta_{\mu_4 \nu_4} 
   \pm (\mbox{the\ other\ 23 terms}) 
  \right] \, , 
 } 
 \label{eq:exp-eps} 
\end{eqnarray} 
where the other 23 terms are obtained 
by shuffling the order of 
$\{ \nu_1, \nu_2, \nu_3, \nu_4 \}$ 
in all possible ways. 
 Each term contributes in the bracket 
with the sign $+$ ($-$) 
if the even (odd) permutation is performed to reach 
this order from $\{ \nu_1, \nu_2, \nu_3, \nu_4 \}$. 
 We found that the method using (\ref{eq:exp-eps}) 
led to a result opposite in sign to the previous result 
(\ref{eq:pi0}). 
 On the other hand, 
REDUCE passed the same test without difficulty.
 Hence, we conclude that it was the case (b) that actually happened. 
 In other words, 
the $\pi^0$ pole contribution in (\ref{eq:pi0}) 
must be changed  
\begin{equation} 
 a_\mu(\pi^0) = 55.60~(3) \times 10^{-11} \, , 
  \label{eq:pi0_new} 
\end{equation} 
in the nVMD model. 

 The sign of the pseudoscalar pole contribution 
 (\ref{eq:pi0_new}) 
has returned to 
that of Ref.~\cite{Kinoshita:1984it} 
and agrees with that of Ref.~\cite{Knecht:2001qf}. 
 But, in the former case, 
the program containing the product of two $\epsilon$-tensors 
has also been used to extract $a_\mu$. 
 At the same time, there was an error in 
the sign of the logarithmic term 
dominating the pseudoscalar pole contribution, 
which was corrected in Ref.~\cite{Hayakawa:1995ps} 
\footnote{ 
 The delicate point will be described in Ref.~\cite{HK}.
}. 
 The double switching of the sign 
has led accidentally to the positive value 
for the pseudoscalar contribution  
in Ref.~\cite{Kinoshita:1984it}. 
\vspace{0,5cm} 

 We conclude that 
the sign of all the results 
for the pseudoscalar pole contribution 
as well as the axial-vector meson pole contribution described in 
Ref.~\cite{Hayakawa:1995ps,Hayakawa:1997rq} 
must be reversed. 
 The signs of the charged pseudoscalar loop contribution 
and the constituent quark loop contribution 
are not affected by the problem noted here. 
 Collecting those results from Eq.~(1.8)  
and changing the sign of Eqs.~(1.9) and (5.1) 
of Ref.~\cite{Hayakawa:1997rq}, 
we obtain the new value 
\begin{equation} 
 a_\mu({\rm LL}) = 89.6~(15.4) \times 10^{-11} \, . 
\end{equation} 
as the current value of the ${\cal O}(\alpha^3)$ 
hadronic light-by-light scattering contribution to 
the muon $g-2$. 
 This reduces the discrepancy between the measurement and 
the prediction of the standard model to 1.6 $\sigma$ deviations. 

\section*{Acknowledgments} 

 The authors would like to thank M.~Knecht and A.~Nyffeler 
for informing their results 
prior to putting their papers on the web 
and for useful communications. 
 They thank A.~Czarnecki 
for letting them know that he also noticed 
a subtlety of the $\epsilon$-tensor in FORM. 
 They thank J.~Vermasseren for a useful communication. 
 M.~H. thanks R.~Kitano, S.~Kiyoura and N.~Yamada 
for their help in setting up his computer 
for his numerical work. 
 The work of T.~K. is supported in part 
by the U.~S.~National Science Foundation. 


\end{document}